\documentclass[aps,prl,twocolumn,reprint,superscriptaddress]{revtex4-1}
\usepackage{graphicx} 
\usepackage{amssymb}
\usepackage{dcolumn}
\usepackage{bm} 
\newcommand{\Tc}{T_{\rm c}}
\newcommand{\dinter}{d_{\rm inter}}
\newcommand{\ddoped}{d_{\rm doped}}
\newcommand{\Hc}{H_{\rm c2}}
\newcommand{\dSC}{d_{\rm sc}}
\newcommand{\dSCa}{d_{\rm sc1}}
\newcommand{\dSCb}{d_{\rm sc2}}

\begin{document}
\title{Tunable Coupling of Two-Dimensional Superconductors in Bilayer SrTiO$_3$ Heterostructures}
\author{Hisashi Inoue}
\affiliation{Geballe Laboratory for Advanced Materials, Department of Applied Physics, Stanford University, Stanford, CA 94305, USA}
\author{Minu Kim}
\affiliation{Stanford Institute for Materials and Energy Sciences, SLAC National Accelerator Laboratory, Menlo Park, CA 94025, USA}
\author{Christopher Bell}
\affiliation{Stanford Institute for Materials and Energy Sciences, SLAC National Accelerator Laboratory, Menlo Park, CA 94025, USA}
\author{Yasuyuki Hikita}
\affiliation{Stanford Institute for Materials and Energy Sciences, SLAC National Accelerator Laboratory, Menlo Park, CA 94025, USA}
\author{Srinivas Raghu}
\affiliation{Stanford Institute for Materials and Energy Sciences, SLAC National Accelerator Laboratory, Menlo Park, CA 94025, USA}
\affiliation{Department of Physics, Stanford University, Stanford, CA 94305, USA}
\author{Harold Y. Hwang}
\affiliation{Geballe Laboratory for Advanced Materials, Department of Applied Physics, Stanford University, Stanford, CA 94305, USA}
\affiliation{Stanford Institute for Materials and Energy Sciences, SLAC National Accelerator Laboratory, Menlo Park, CA 94025, USA}

\date{\today}
\begin{abstract}
Interlayer coupling effects between high mobility two-dimensional superconductors are studied in bilayer $\delta$-doped SrTiO$_3$ heterostructures. By tuning the undoped SrTiO$_3$ spacer layer between the dopant planes, clear tunable coupling is demonstrated in the variation of the sheet carrier density, Hall mobility, superconducting transition temperature, and the temperature- and angle-dependences of the superconducting upper critical field. Systematic variation is found between one effective (merged) two-dimensional superconductor to two decoupled two-dimensional superconductors. In the intermediate coupled regime, a crossover is observed due to coupling arising from inter-sub-band interactions.
\end{abstract}
\maketitle

Interlayer coupling between two-dimensional (2D) planes in layered materials is a key parameter which dramatically influences their physics and functionality. Paradigmatic examples in the field of superconductivity include the high temperature superconducting cuprates \cite{Anderson1995, Tsvetkov1998}, and the iron pnictides \cite{Xu2011}, as well as artificially created superlattices, in which the Josephson coupling between the layers can be tunable continuously using thin film control \cite{Jin1989,Blatter1994}. There have also been extensive studies of coupled bilayer 2D systems, where exotic phases arise from the interlayer interactions \cite{Gramila1991,Eisenstein1992,Eisenstein2004,Tiemann2008a,Geim2007,Gorbachev2012,Maher2013}.

Among the various 2D systems currently under investigation, SrTiO$_3$ (STO) is of particular interest since it is a rare example of a high mobility doped semiconductor that is simultaneously superconducting. By confining electrons in STO two-dimensionally using field effect gating \cite{ueno_natmat2008}, heterointerfaces with LaTiO$_3$ \cite{biscarasPRL}, or LaAlO$_3$ \cite{ohtomo_nature2004,reyren_science2007}, and $\delta$-doping \cite{Kozuka2009}, the interplay between sub-band quantization and superconductivity can be investigated. Here we exploit interlayer coupling in a $\delta$-doped bilayer system to {\it spatially} and flexibly control the sub-bands. By systematically tuning the interlayer coupling by varying the spacer layer thickness, we can examine the effects of the sub-band quantization on superconductivity. 

\begin{figure}[t]
\includegraphics{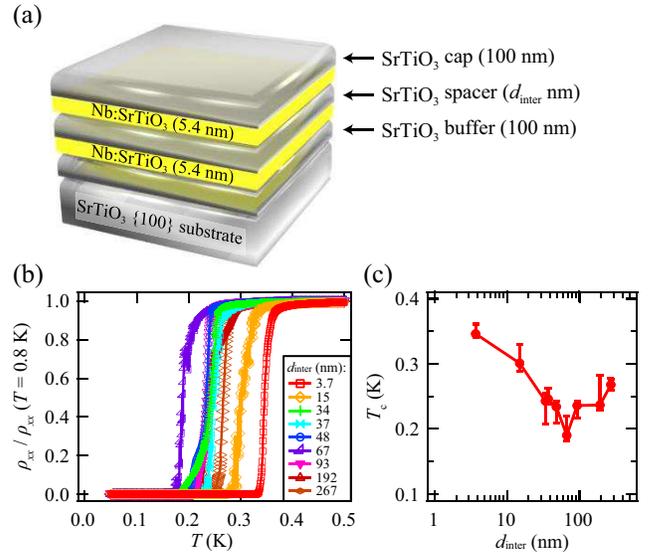}
\caption{\label{fig1}(color online) (a) Schematic diagram of the STO bilayer $\delta$-doped structure.  (b) Temperature dependence of $\rho_{xx}$ for various $\dinter$. Data are normalized by the sheet resistance at $T$ = 800 mK. (c) Variation of $\Tc$ with $\dinter$. Temperature error bars correspond to the 90 \% - 10 \% transition width.}
\end{figure}
The bilayer samples consist of two Nb:STO $\delta$-doped layers, with an undoped STO spacer of thickness $\dinter$, as well as 100 nm thick undoped STO cap and buffer layers. All samples were fabricated using pulsed laser deposition using growth conditions as reported elsewhere \cite{Kozuka2010a}. The thickness of each of the two $\delta$-layers was fixed at a constant value of $\ddoped = 5.4 \pm 1$ nm, as calibrated from the total thickness and the laser pulse count. The interlayer undoped STO thickness $\dinter$ was varied in the range $3.7 \le \dinter \le 272$ nm. The dopant concentration of the $\delta$-doped layers was 1 at.\%. A schematic of the bilayer samples is shown in Fig$.$ \ref{fig1} (a). Electronic transport measurements were performed in a helium cryostat at $T = 2$ K and in a dilution refrigerator from $T = 800$ mK down to $T = 50$ mK. The latter system was fitted with an {\it in-situ} horizontal rotator, enabling the inclination angle between the applied magnetic field and the sample normal to be varied. 

\begin{figure}[t]
\includegraphics{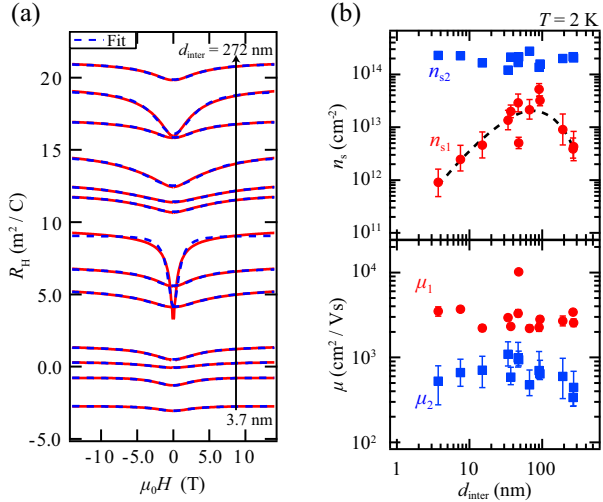}
\caption{\label{fig2}(color online) (a) Magnetic field dependence of the Hall co-efficient $R_{\rm H}$ at $T = 2$ K. Data are offset vertically for clarity. Solid curves are the experimental data. Dashed lines are the best fits to the two-carrier model. (b) Variation of the sheet carrier density and Hall mobility with $\dinter$ for each component extracted from the two-carrier fit in (a). Circles (squares) correspond to the relatively high (low) mobility component $\mu_1$ ($\mu_2$), and the low (high) sheet carrier density component $n_{\rm s1}$ ($n_{\rm s2}$), respectively. Dashed curve is a guide to eye. Error bars for $\mu_1$ and $n_{\rm s2}$ are smaller than the point size.}
\end{figure}
For the various $\dinter$, superconductivity was found with transition temperature $\Tc$ ranging from 190 to 346 mK, as shown in Fig$.$ \ref{fig1}(b). Here $\Tc$ is defined as the temperature where longitudinal sheet resistance $\rho_{xx}$ drops to half the value of $\rho_{xx}$ at $T = 800$ mK. The data shows a clear modulation of $\Tc$, which drops for intermediate $\dinter$ [Fig$.$ \ref{fig1}(c)] already suggestive of variable coupling between the two superconducting $\delta$-layers through the undoped STO. The normal state characteristics were measured by Hall effect up to a magnetic field of $\mu_0 H = 14$ T at $T = 2$ K ($\mu_0$ is the vacuum permeability). A non-linearity in the Hall effect data was observed, as shown clearly in the Hall co-efficient, $R_{\rm H}$,  $vs.$ magnetic field characteristics in Fig$.$ \ref{fig2}(a). This indicates that there are at least two types of carriers with different mobilities and carrier densities contributing to the conduction in parallel. As shown in the figure, these data could be well parameterized using the well-known two-carrier form, assuming the existence of two types of electrons with mobilities $\mu_1$, $\mu_2$ and sheet carrier densities $n_{\rm s1}$, $n_{\rm s2}$. Here the fitting parameters are $n_{\rm s1}$, $n_{\rm s2}$ and $\mu_1$, using $\rho_{xx}$ at zero magnetic field as a constraint. 

The best fit values extracted for $\mu_1$, $\mu_2$, $n_{\rm s1}$ and $n_{\rm s2}$ are shown in Fig$.$ \ref{fig2}(b), with the relatively high mobility, low sheet carrier density data points ($\mu_1$, $n_{\rm s1}$) shown as circles, and the lower mobility, higher sheet carrier density data ($\mu_2$, $n_{\rm s2}$) shown as squares. A clear enhancement of $n_{\rm s1}$ by around an order of magnitude is evident at the intermediate interlayer thickness range, co-incident with the suppression of the $\Tc$. These data strongly suggest a change in the interlayer coupling as $\dinter$ is varied, and the presence of relatively high mobility conduction electrons in the undoped interlayer region when $\dinter \sim 67$ nm. We note that the $\Tc$ is suppressed when $n_{\rm s1}$ is enhanced, which is distinct to recent studies that have suggested high mobility carriers are essential for superconductivity in asymmetrically confined STO electron gases \cite{biscarasPRL}.
\begin{figure}[b]
\includegraphics{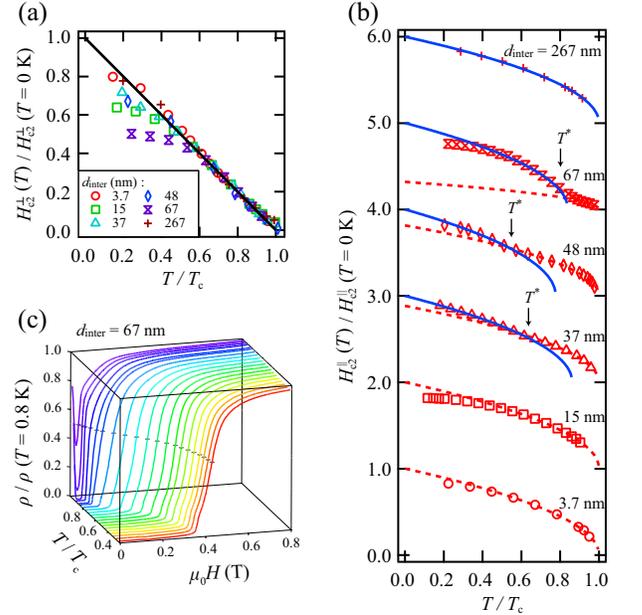}
\caption{\label{fig3}(color online) (a) Temperature dependence of the out-of-plane upper critical field $\Hc^{\perp}$ for various $\dinter$. Solid line is the fit to the GL equation. (b) Temperature dependence of $\Hc^{\parallel}$ for various $\dinter$. Data are offset vertically for clarity. The data were fitted to Eq$.$ (\ref{eq_Hc2_para}) for relatively low and high temperatures where best fits are shown in solid and dashed lines. $T^*$ shows the crossover temperatures where the dominant length scale changes on $\Hc^{\parallel}$. (c) Magnetic field dependence of $\rho_{xx}$ at various temperatures for $\dinter$ = 67 nm sample. Markers show the 50 \% criterion.}
\end{figure}

\begin{figure*}[t]
\includegraphics{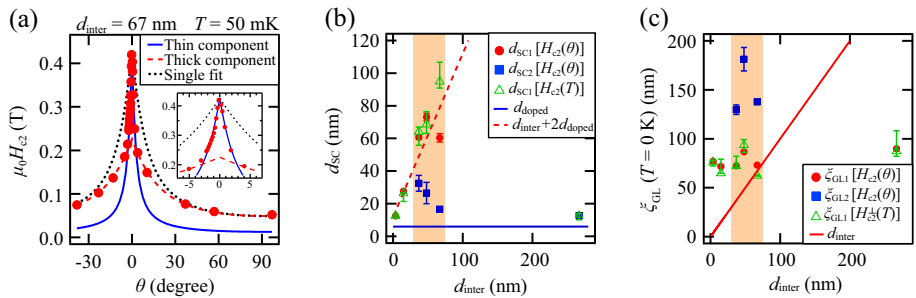}
\caption{\label{fig4}(color online) (a) Angular dependence of $\Hc$ for $\dinter$ = 67 nm. $\theta = 0^\circ$ is defined as the direction parallel to the sample plane. Data for angles around $\theta = 0^\circ$ and $\theta = 90^\circ$ were separately fitted by Eq$.$ (\ref{eq_Hc2_angle}). The best fits are shown as solid and dashed curves respectively. For comparison, a single fit using Eq$.$ (\ref{eq_Hc2_angle}) (plotted as a dotted line) cannot describe the data for whole angle range. Inset: The range around $\theta = 0^{\circ}$ is enlarged for the same data. (b) Variation of $\dSC$ and (c) $\xi_{\rm GL}$ with $\dinter$. Data shown in closed and open markers are estimated from $\Hc (\theta )$ and $\Hc (T)$ respectively. Lines correspond to the nominal thicknesses of the layers. $\xi_{\rm GL1}$ and $\xi_{\rm GL2}$ are $\xi_{\rm GL}$ for the thicker and thinner components respectively. Shaded areas in (b) and (c) represent the intermediate $\dinter$ range where similar features were observed in $H_{\rm c2} (T)$ and $H_{\rm c2} (\theta)$ data as for $\dinter = 67$ nm. }
\end{figure*}
The superconducting properties were studied in detail by examining the temperature-dependence of the superconducting upper critical field $\Hc$ in the in-plane and out-of-plane configurations. All of the samples showed the expected linear $T$-dependence of the out-of-plane upper critical fields $\Hc^\perp$, as shown in Fig$.$ \ref{fig3}(a) following the standard linearized Ginzburg-Landau (GL) theory for superconductors,
$\Hc^{\perp}(T) = {\it \Phi}_0(1 - T / \Tc )(2 \pi \xi_{\rm{GL}}(0)^2)^{-1}$, where ${\it \Phi}_0$ is the flux quantum and $\xi_{\rm GL}(0)$ is the GL coherence length $\xi_{\rm GL}$ at $T=0$ K. The in-plane $\Hc$ data, $\Hc^{\parallel}(T)$ showed rather more complex behavior, as shown in Fig$.$ \ref{fig3}(b). As for the $\Hc^{\perp}(T)$ data, here $\Hc^{\parallel}(T)$ was extracted from $\rho_{xx}(H)$ using the same definition as for $\Tc$, as shown by the markers in the example raw data in Fig$.$ \ref{fig3}(c). In the relatively thin and thick $\dinter$ regimes, the $\Hc^{\parallel}(T)$ data can be well-fitted to the 2D GL form,
\begin{equation}
\label{eq_Hc2_para}
\Hc^{\parallel} \left( T \right) = \frac{ {\it \Phi}_0 \sqrt{12} }{ 2 \pi \xi_{\rm{GL}}(0) d_{\rm SC} } \Bigl( 1 - \frac{ T }{ T_{\rm c} } \Bigr) ^{\frac{1}{2}},
\end{equation}
where $d_{\rm SC}$ is the length scale of the spatial distribution of the order parameter. The key result here is that intermediate thickness samples ($d_{\rm inter}$ = $37 \sim 67$ nm) exhibit a clear upturn in $\Hc^{\parallel}(T)$  at temperatures below some characteristic $T^*$. Since the form of $\Hc^{\parallel}(T)$ is still 2D-like both above and below $T^*$, these data were fit to Eq$.$ (\ref{eq_Hc2_para}) using two different $\Tc$ and $\Hc^{\parallel}(0)$ above and below $T^*$, to give a quantitative parameterization. The fits showed good agreement with the experimental data, implying the existence of two different superconducting length scales depending on temperature and magnetic field in the same system. 

The field inclination angle-dependence, $\Hc (\theta)$ was also measured, as shown in Fig$.$ \ref{fig4}(a) for the case $\dinter$ = 67 nm. In 2D superconductors, $\Hc (\theta)$ is strongly anisotropic since the effectiveness of the orbital pair breaking is dependent on the magnetic field direction. In analogy with the $T^*$ found in the $T$-dependence, there is a characteristic kink in the $\Hc (\theta)$ data here. Similar features were also observed for $\dinter = 37$ nm and 48 nm. Again we can interpret these data by fitting them to two separate 2D forms, using the $\Hc (\theta)$ form first derived by Tinkham \cite{Tinkham1963}: 
\begin{equation}
\label{eq_Hc2_angle}
\Bigl|\frac{ H_{\rm c2} \left( \theta \right) \sin \theta}{ H_{\rm c2}^\perp } \Bigr| + \Bigl( \frac{ H_{\rm c2} \left( \theta \right) \cos\theta }{ H_{\rm c2}^{\parallel} } \Bigr)^2 = 1.
\end{equation}
Here we note that the data could not be described by a fit using the single form of Eq$.$ (\ref{eq_Hc2_angle}) plotted as a dotted line in Fig$.$ \ref{fig4}(a). From a fit to Eq$.$ (\ref{eq_Hc2_angle}) we derive $\dSC$ and $\xi_{\rm GL}$, as shown in Fig$.$ \ref{fig4}(b) and (c). These parameters were also estimated from the temperature dependent critical field data, leading to similar values. 

The form of $\dSC$ as a function of $\dinter$, shown in Fig$.$ \ref{fig4}(b) is of particular interest. The thicker $\dSC$ component, denoted hereafter as $\dSCa$ which is dominant near $\theta$ = 90$^\circ$, initially follows the total layer thickness $d_{\rm tot}$ = $\dinter + 2\ddoped$ but disappears in the thick limit. In contrast, the thinner $\dSC$ component, $\dSCb$ dominant near $\theta$ = 0$^\circ$, is always of the order of $\ddoped$. As $\dinter$ increases, the system shows a crossover from the coupled state where the two doped layers are strongly linked and behave as a single 2D superconductor, to a fully decoupled state where they behave as two independent 2D layers. As emphasized in Fig$.$ \ref{fig3}(b) and Fig$.$ \ref{fig4}(a), we can see a rather similar transition by either decreasing the temperature below $\Tc$ or rotating the magnetic field at fixed $T$.

We discuss these results in the context of previous studies in metallic superconductor/insulator/superconductor (S/I/S) heterostructures. Here we note that we assume an isotropic $\xi_{\rm GL}$, reflecting the bulk isotropy of STO, and neglect any possible influence of the symmetry breaking confining electric fields in the direction perpendicular to the layers. In the S/I/S case, the two S layers are Josephson coupled across the I layer when the insulating layer thickness $\dinter$ is thin compared to the GL coherence length in perpendicular direction $\xi_{\rm GL}^\perp$,  while it is suppressed for thick $\dinter$ regime. The characteristic forms of $\Hc^{\parallel}(T)$ in the intermediate $\dinter$ regime can also be understood in this model: since $\xi_{\rm GL}$ is an increasing function of $T$, the S layers are decoupled for $T < T^*$, where $T^*$ satisfies $\xi_{\rm GL}(T^*) \sim \dinter/\sqrt{2}$ \cite{Klemm1975,Takahashi1986,Jin1989,Blatter1994}. Consequently the coupled/decoupled transition at $T^*$ results in a kink in the $\Hc^{\parallel}(T)$ data. While this qualitatively describes the results, experimentally we found that $\xi_{\rm GL} (T^*)$ is always larger than $\dinter / \sqrt{2}$ in the samples which show kinks in the $H_{\rm c2} (T)$ data, since the $\xi_{\rm GL}$ at $T$ = 0 K is similar to or greater than $\dinter$ as shown in Fig$.$ \ref{fig4}(c).

\begin{figure}[t]
\includegraphics{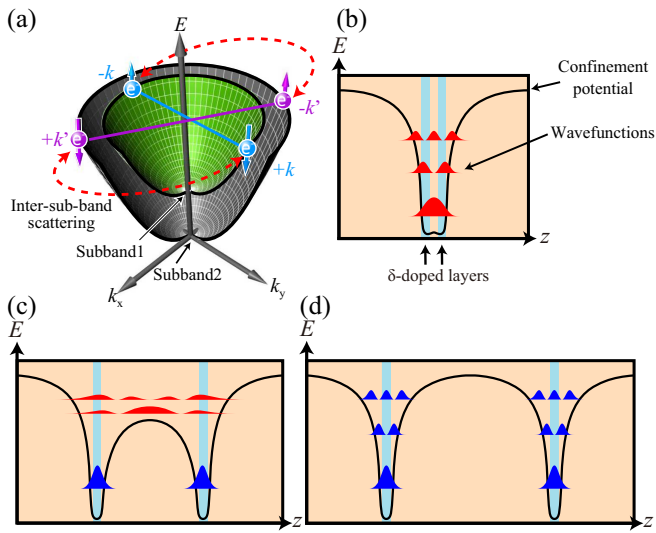}
\caption{\label{fig5}(color online) (a) Schematic drawing of Cooper pairs formed on different sub-bands. The parabolic surfaces represent the dispersion relation of the normal state electrons in momentum space. Arrows with dashed lines represents the inter-sub-band interaction of Cooper pairs at the Fermi level. (b)-(d) Schematic sub-band structure of the STO $\delta$-doped bilayers for relatively small, intermediate and large $\dinter$ respectively. $E$ is the electron energy. $z$ is the coordinate perpendicular to the layers. The solid curves and filled areas represent the confinement potential and the probability amplitude of the envelope electron wavefunctions respectively. Nominally areas of $\delta$-doped planes are shown as shaded blocks. In reality more sub-bands could be occupied than shown here. }
\end{figure} 
In addition to this quantitative discrepancy, the enhancement of the high mobility carrier density $n_{\rm s1}$ as well as modulation of $\Tc$ in the intermediate $\dinter$ range suggests that the simple S/I/S concept may not be applicable to this semiconductor system. We emphasize that many experimental and theoretical studies of this and related 2D electron systems in STO suggest the presence of quantized sub-bands \cite{Popovic2005,Popovic2008,Son2009,Kozuka2009,Jalan2010,BenShalom2010,caviglia2010,Santander-Syro2011,Meevasana2011,Delugas2011,Kim2011,Khalsa2012}. Thus a consistent understanding of both the normal and the superconducting states requires consideration of the sub-band structure in the electrostatic confinement potential which must be considered self-consistently throughout the system as a whole. In such a system it is possible that each sub-band has different superconductivity characteristics, $e.g.$ superconducting gap, pairing potential, $\xi_{\rm GL}$ etc. However inter-sub-band scattering should be taken into account, since although the bottom of the sub-bands can be separated by energies significantly larger than the superconducting gap, at the Fermi level Cooper pairs can be scattered into different bands since momentum and energy are both conserved [Fig$.$ \ref{fig5}(a)].

From the considerations above, next we discuss the modulation of sub-band structure as the source for the observed features, in particular noting that sub-bands can be generated in the interlayer spacer itself. In the thin $\dinter$ limit, many of the sub-bands are spread across the interlayer, as sketched schematically in Fig$.$ \ref{fig5}(b). Here most of the sub-bands show relatively low mobility due to scattering by the Nb dopants, and $n_{\rm s1}$ is relatively small. In the superconducting state, the condensate shows thickness $\sim 2\ddoped + \dinter$ reflecting the normal state electron distribution. In the intermediate $\dinter$ regime, Fig$.$ \ref{fig5}(c), the cleaner undoped interlayer is thick enough that the high mobility electron density $n_{\rm s1}$ in the upper sub-bands is enhanced. In the superconducting state the coupling between the two superconducting layers through the Cooper pair channel in the upper sub-bands in the interlayer create a superconducting layer with spatial extent $\dSCa \sim 2\ddoped + \dinter$ in low magnetic fields. In higher magnetic fields, however, orbital limit pair breaking occurs first in the upper sub-bands. This decouples the two narrow condensates confined around the $\delta$-doped layers with the thickness $\dSCb \sim \ddoped$. This manifests as the characteristic features in $\Hc^{\parallel}(T)$ and $\Hc(\theta)$ in Figs$.$ \ref{fig3}(b) and \ref{fig4}(a). When $\dinter$ is increased further, the system behaves as two decoupled 2D superconductors, with no significant carrier weight in the interlayer region, leading to again relatively low $n_{\rm s1}$, and a superconducting thickness $\sim \ddoped$, as shown in Fig$.$ \ref{fig5}(d).

The modulation of $\Tc$ may be related to the variable interaction between the upper and the lower sub-bands: in the thick and thin $\dinter$ regime, the superconductivity is robust due to the efficient inter-sub-band coupling from the large spatial overlap between the upper and the lower sub-bands. In contrast for the intermediate $\dinter$, superconductivity is weakened since the spatial overlap of the upper and the lower sub-bands is reduced, leading to a suppression of average $\Tc$. In addition, given the possibility that the orbital character and superconducting pairing strengths may also be dependent on the sub-band index in these heterostructures, a more thorough theoretical analysis, extending the work of single layer systems \cite{Fernandes2013,Mizohata2013}, is motivated.

In conclusion, we investigated the interlayer coupling effects between two 2D high mobility superconducting wells. By systematically changing the interlayer thickness, we observed a coupled to decoupled crossover in both normal and superconducting state driven by the modulation of sub-band structure. We emphasize the flexibility of this composite structure, which enables to reach an interesting regime where the inter-sub-band interaction may be dominant mechanism for interlayer coupling. The current structures offer a model system to control the bilayer degree of freedom in multi-component superconductors, which has been theoretically proposed to synthesize unconventional superconductivity \cite{Sato2010,Fu2010,Nakosai2012}.

This work is supported by the Department of Energy, Office of Basic Energy Sciences, Division of Materials Sciences and Engineering, under Contract No. DE-AC02-76SF00515. Discussions with M. R. Beasley and K. Moler are acknowledged.

\end{document}